\documentclass[aps, prl, twocolumn, 10pt, superscriptaddress, floatfix]{revtex4-1}

\usepackage{graphicx}        
\usepackage{epstopdf}
\usepackage{CJK}
\usepackage{bm}
\usepackage{amssymb}
\usepackage{amsmath}
\usepackage{siunitx}
\usepackage{pifont}
\usepackage[usenames,dvipsnames]{xcolor}
\usepackage[linkcolor=violet,citecolor=blue,colorlinks=true,urlcolor=purple]{hyperref}

\def\cep{\textrm{CEP}}

\newcommand{\B}[1]{\textbf{#1}}

\newcommand{\SM}{{Supplemental Material \cite{SMHanus2019}}}

\renewcommand{\vec}{\textbf}

\begin{document}
\begin{CJK*}{UTF8}{gbsn}

\title{Experimental Separation of Sub-Cycle Ionization Bursts in Strong-Field Double Ionization of H$_2$}

\author{V\'{a}clav\,Hanus}
\email[]{vaclav.hanus@tuwien.ac.at}
\affiliation{Photonics Institute, Technische Universit\"at Wien, A-1040 Vienna, Austria}

\author{Sarayoo\,Kangaparambil}
\affiliation{Photonics Institute, Technische Universit\"at Wien, A-1040 Vienna, Austria}

\author{Seyedreza\,Larimian}
\affiliation{Photonics Institute, Technische Universit\"at Wien, A-1040 Vienna, Austria}

\author{Martin\,Dorner-Kirchner}
\affiliation{Photonics Institute, Technische Universit\"at Wien, A-1040 Vienna, Austria}

\author{Xinhua\,Xie (谢新华)}
\affiliation{Photonics Institute, Technische Universit\"at Wien, A-1040 Vienna, Austria}
\affiliation{SwissFEL, Paul Scherrer Institute, 5232 Villigen PSI, Switzerland}

\author{Markus\,S.\,Sch\"{o}ffler}
\affiliation{Institut f\"ur Kernphysik, Goethe-Universit\"at, D-60438 Frankfurt am Main, Germany}

\author{Gerhard\,G.\,Paulus}
\affiliation{Institute for Optics and Quantum Electronics, Friedrich-Schiller-Universit\"at Jena, D-07743 Jena, Germany}

\author{Andrius\,Baltu\v{s}ka}
\affiliation{Photonics Institute, Technische Universit\"at Wien, A-1040 Vienna, Austria}

\author{Andr\'{e} Staudte}
\affiliation{Joint Attosecond Science Lab of the National Research Council and the University of Ottawa, Ottawa, Canada}

\author{Markus\,Kitzler-Zeiler}
\email[]{markus.kitzler@tuwien.ac.at}
\affiliation{Photonics Institute, Technische Universit\"at Wien, A-1040 Vienna, Austria}

\begin{abstract}
We report on the unambiguous observation of the sub-cycle ionization bursts in sequential strong-field double ionization of H$_2$ 
and their disentanglement in molecular frame photoelectron angular distributions.
This observation was made possible by the use of few-cycle laser pulses with a known carrier-envelope phase in combination with multi-particle coincidence momentum imaging.
The approach demonstrated here will allow
sampling of the intramolecular electron dynamics and the investigation of charge-state specific Coulomb-distortions on emitted electrons in polyatomic molecules.
\end{abstract}





\maketitle
\end{CJK*}

The creation of free or bound electron wave packets with a sub-laser-cycle
temporal structure
using intense ultrashort laser pulses is at the heart of most processes in attosecond physics \cite{Kienberger2002, Xu2010b, Haertelt2016, Xie2012_interferometry}. The strong oscillating electric field of the laser light can  cause ionization of a system via half-cycle periodic emissions of electron wavepackets and imprints a dominant sub-cycle temporal structure onto them. In addition, these periodically emitted wavepackets, known as the sub-cycle ionization bursts, may be modulated by the non-adiabatic response of the strongly driven system \cite{Takemoto2010, Odenweller2011, Xie2012_interferometry, Spanner2012}.
In strong-field double ionization the sub-cycle ionization bursts from the first ionization step can be correlated with the ionization bursts from the second ionization step. Since each ionization burst corresponds to a specific time within the laser pulse envelope, various delays between the two ionization steps are sampled within a single pulse.
In molecules, the delay between the two ionization steps can determine the outcome of molecular fragmentation reactions \cite{Xie2015, Sandor2016}, constitutes the probe delay for intramolecular electron dynamics \cite{Takemoto2010, Spanner2012, Hennig2005}, and determines the molecular site from where the photoelectron originates \cite{Ohmura2006b, Wu2012b, Liu2017}.
To extract the ultrafast dynamics underlying these processes from experiments it is necessary to disentangle the separate contributions of the sub-cycle ionization bursts in measured photoelectron or photoion distributions.
However, so far such disentanglement has been achieved only for atomic systems in both the sequential \cite{Schoffler2016} and nonsequential \cite{Bergues2012} regimes of double ionization.

In this Letter, we demonstrate a method that allows for the unambiguous identification and extraction of the contributions 
from specific sub-cycle ionization bursts in electron momentum distributions measured for molecular double ionization.
Using the example of the H$_2$ molecule we show that we can measure the molecular frame photoelectron angular distributions (MF-PADs) for a specific pair of two time-ordered sub-cycle ionization bursts in the sequential double ionization (SDI) process $\textrm{H}_2 \rightarrow \textrm{H}_2^+ + e^- \rightarrow \textrm{H}^+ + \textrm{H}^+ + 2e^-$.
The  observation of the sub-cycle bursts and the unambiguous assignment of the two detected electrons to the first and second ionization burst was accomplished
by combining our technique developed for atoms \cite{Schoffler2016}
with our recently published method for the reconstruction of molecular dynamics \textit{during} laser interaction \cite{Hanus2019}.

\begin{figure*}[tb]
  \begin{center}
  \includegraphics[width=0.99\textwidth]{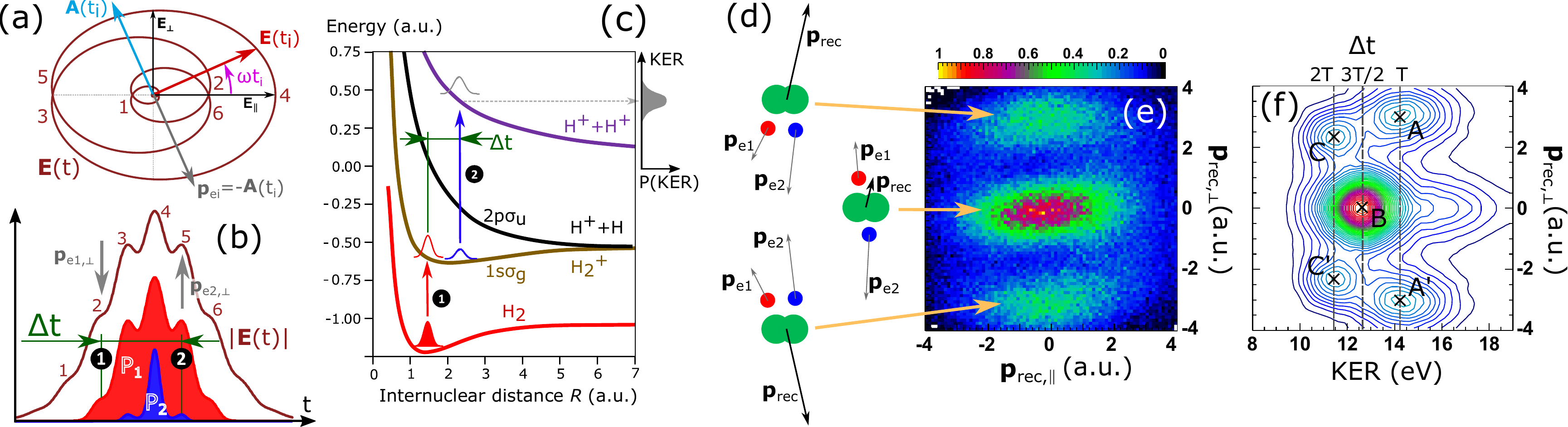}
  \end{center}
  \caption{(a) Electric field $\vec{E}(t)$ in the polarization plane of a pulse with $\cep=0$. The vectors of the electric field $\vec{E}(t_i)$ and vector potential $\vec{A}(t_i)$ at the instant of the emission $t_i$ of an electron with momentum $\vec{p}_{ei}$ are indicated.
  (b) Absolute value of the field evolution shown in (a), $|\vec{E}(t)|$. The red and blue areas, denoted $P_1$ and $P_2$, sketch the ionization rates of the first respectively second ionization step with the sub-cycle ionization bursts at the peaks of $|\vec{E}(t)|$ (numbered 1--6).  Two ionization instants delayed by $\Delta t=1.5T$ with each other, marked by \ding{182} and \ding{183}, and the corresponding dominant directions of the $\perp$-momenta of the two emitted electrons, $\vec{p}_{e1,\perp}$ and $\vec{p}_{e2,\perp}$, are indicated.
  (c) Potential energy curves of H$_2$ relevant for SDI with a delay $\Delta t$ between the two ionization steps \ding{182} and \ding{183}.
  (d) Sketches of momentum vectors for three cases of SDI dynamics corresponding to the three lobes in (e), as indicated by arrows.
  (e) Measured recoil momentum vector $\vec{p}_\textrm{rec} = \vec{p}_{\textrm{H}^+} + \vec{p}_{\textrm{H}^+} = - (\vec{p}_{e1} + \vec{p}_{e2})$ in the laser polarization plane perpendicular $\perp$ and parallel $\parallel$ to the main axis of the polarization ellipse, cf. (a).
  (f) $\vec{p}_{\textrm{rec},\perp}$ vs. KER. The upper and lower lobes from (e) appear as separated peaks \B{A/C} and \B{A'/C'}, respectively, the center lobe appears as peak \B{B}.  }
  \label{fig1}
\end{figure*}


In our experiments, we used a reaction microscope \citep{Doerner2000} to measure in coincidence the momenta of two protons and two electrons created upon interaction of a cold jet of H$_2 $ with intense, elliptically polarized few-cycle laser pulses. 
The ellipticity, defined as the ratio of the electric field strength perpendicular and parallel to the main axis of the polarization ellipse, was $E_\perp/E_\parallel=0.85$, cf. Fig.~\ref{fig1}(a).
The laser center wavelength was $\lambda=750$\,nm. Thus, the laser oscillation period $T=2\pi/\omega=2.5$\,fs with $\omega=2\pi c/\lambda$ the frequency and $c$ the speed of light.
The duration of the pulses ($\approx$\SI{4.5}{\fs}) and their carrier-envelope phases (CEPs) were measured with a stereo electron spectrometer in phase-tagging mode \citep{Rathje2012}.
The laser beam was focused in an ultra-high vacuum chamber (base pressure 10$^{-10}$\,mbar) onto a supersonic gas jet of H$_2$. 
Ions and electrons emerging from the interaction volume were guided by weak electric (21\,V/cm) and magnetic fields (12\,G) to two position-sensitive detectors.
%
The momentum of the second electron was calculated from the momenta of the measured electron and both ions by exploiting momentum conservation. 
%
The laser peak intensity \textit{in situ} was \SI{9e14}{\watt\per\square\cm} \citep{Smeenk2011}.
Further details on the reaction microscope can be found in Refs.\,\cite{Xie2012_CE, Zhang2014a, Xie2017b} and on the optical setup in Ref.\,\cite{Schoffler2016}. 

When $\textrm{H}_2$ undergoes SDI in a strong elliptically polarized laser field, the first ionization step at time $t_1$ triggers vibronic dynamics in the molecular cation $\textrm{H}_2^+$, and the second ionization step at time $t_2=t_1+\Delta t$ initiates Coulomb explosion into $\textrm{H}^+ + \textrm{H}^+$, cf. the sketches in Figs.~\ref{fig1}(b) and \ref{fig1}(c). 
While the kinetic energy released, $\textrm{KER}=1/R$, with $R$ the internuclear distance at which the Coulomb explosion is initiated, provides a precise measure for the nuclear motion in between the two ionization steps \cite{Niikura2003, Ergler2006_D2_nuclear_wavepacket, Hanus2019}, the momenta of the two emitted electrons, $\vec{p}_{e1,e2}$, provide information on the emission times of the two electrons within a laser cycle.
This is because in elliptically polarized light
the ionization phase within a laser cycle, $\omega t_i$, is mapped onto the emission angle of the photoelectron via the relation $\vec{p}_{ei} = -\vec{A}(t_i), \: i=\{1,2\}$  \cite{Faisal1973, Reiss1980}, cf. Fig.~\ref{fig1}(a). The laser vector potential $\vec{A}(t)$ is connected to the laser electric field by $\vec{A}(t)=-\int_{-\infty}^{t}\vec{E}(t')dt'$. Thus, measurement of the electron emission angle in the laboratory frame determines the ionization time $t_i$ within one cycle \citep{Maharjan2005, Eckle2008a, Schoffler2016, Hanus2019}.

In the case of SDI, the momenta of the two emitted electrons are reflected in the recoil momentum vector of the two protons, $\vec{p}_\textrm{rec} = \vec{p}_{\textrm{H}^+} + \vec{p}_{\textrm{H}^+} = - (\vec{p}_{e1} + \vec{p}_{e2})$ \cite{Schoffler2016, Pfeiffer2011_Nature_Phys, Maharjan2005}.
Fig.~\ref{fig1}(e) shows the measured distribution of $\vec{p}_\textrm{rec}$ in the polarization plane, integrated over all values of the CEP and KER.
In elliptical light, ionization takes place preferentially around the times when the field vector $\vec{E}(t)$ passes  the major axis of the polarization ellipse during its rotation, i.e., twice during the optical cycle. 
Depending on the ionization delay $\Delta t$, the two electrons can be streaked by the laser field into the same or into opposite hemispheres.
If the two electrons are ejected with a delay of an even number of half cycles [$\Delta t = 2n\frac{T}{2}; n=0,1,2,\hdots$], the electrons are emitted into the same hemisphere and their momenta add up to a large value of $\vec{p}_\textrm{rec}$.
Examples of electron momentum vectors consistent with such SDI dynamics are indicated as the upper and lower sketches in Fig.~\ref{fig1}(d).
%
%
In contrast, if the two electrons are emitted with an odd number of half cycles [$\Delta t = (2n+1)\frac{T}{2}, \; n=0,1,2,\hdots$], they are streaked into the opposite hemisphere and their momenta cancel to a small value of $\vec{p}_\textrm{rec}$, see the center sketch in Fig.~\ref{fig1}(d).

The three lobes in the momentum distribution in Fig.~\ref{fig1}(e) are, thus, the signatures of the sub-cycle ionization bursts emitted during various combinations of half-cycle peaks of $|\vec{E}(t)|$, numbered 1--6 in Fig.~\ref{fig1}(b), integrated over the whole pulse duration and over all values of the CEP and KER.
Because the ionization delay $\Delta t$ is mapped onto KER, cf. Fig.~\ref{fig1}(c), depending on their ionization delay $\Delta t = n\frac{T}{2}, \; n=0,1,2,\hdots$, a certain pair of sub-cycle bursts leads to a certain value of KER.
The three lobes in the momentum distribution Fig.~\ref{fig1}(e), thus, appear as separated peaks \B{A/A', B, C/C'}
in the KER vs. $\vec{p}_{\textrm{rec},\perp}$ distribution shown in Fig.~\ref{fig1}(f).
Hence, selection of a certain range in KER allows selecting a well-defined range of $\Delta t$ values \cite{Hanus2019}.

\begin{figure}[t!]
  \centering
  \includegraphics[width=0.99\columnwidth]{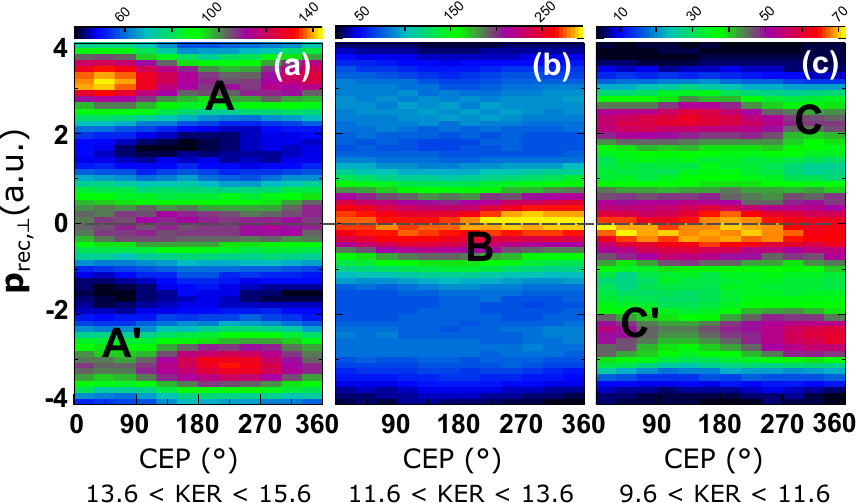}
  \caption{Measured distributions of the recoil momentum vector along the minor polarization axis, $\vec{p}_{\textrm{rec},\perp}$, over CEP for the three indicated ranges of KER (in eV).}
  \label{fig2}
\end{figure}

To understand how the peaks \B{A/A', B, C/C'} can be disentangled into the different contributions of specific SDI bursts, we first resolve them as a function of CEP for three different ranges of KER, each 2\,eV wide, corresponding to about $0.5T$-wide ranges of $\Delta t$ around $1T, 1.5T$ and $2T$, see Figs.~\ref{fig2}(a)-(c).
A pronounced dependence on the CEP of both the momentum and yield is clearly visible for all peaks. 
Particularly strong yield modulations are observed for peaks \B{A/A'} and \B{C/C'}, where for certain values of the CEP the two electrons are predominantly emitted into one hemisphere only. Additional analysis of this asymmetric two-electron emission and its application to the calibration of the CEP in the experimental distributions is provided in the \SM{}.

\begin{figure}[t!]
  \centering
  \includegraphics[width=0.99\columnwidth]{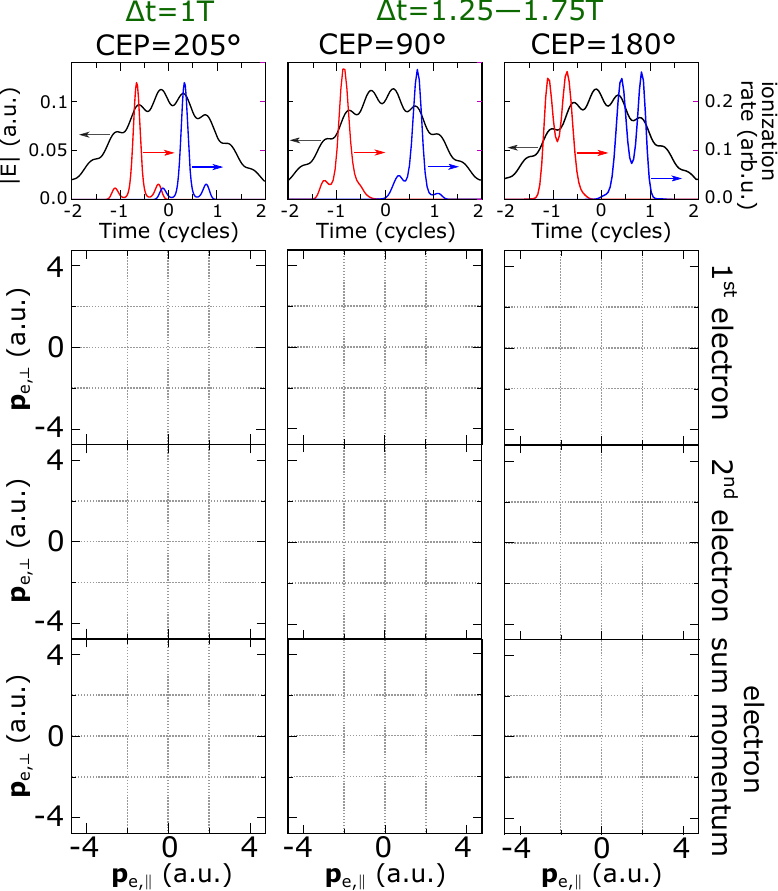}
  \caption{
  Top row: absolute value of electric field, $|\vec{E}(t)|$,  (black) and simulated SDI rates (red, blue) over time for different values of the $\cep$ and $\Delta t$ (columns).
  Bottom rows: corresponding simulated electron momentum distributions in the polarization plane. See text for details.}
  \label{fig3}
\end{figure}

A qualitative understanding of the CEP-dependence of the SDI dynamics can be obtained with the help of semi-classical simulations.
The numerical model underlying these simulations is described in the \SM{}, results are shown in Fig.~\ref{fig3}.
First, we apply our model to the measured momentum distributions for the highest KER-range, shown in Fig.~\ref{fig2}(a), for which $\Delta t \approx 1T$.
For this ionization delay the two electrons are both streaked into the same hemisphere, either both upwards or both downwards, depending on the specific combination of sub-cycle bursts.
The strong yield-asymmetry and the dominance of the lower peak \B{A'} in Fig.~\ref{fig2}(a) for $\cep=205^\circ$ shows, because of $\vec{p}_{\textrm{rec},\perp} = - (\vec{p}_{e1,\perp} + \vec{p}_{e2,\perp})$, that for this CEP-value the two electrons are predominantly emitted only into the {upwards} hemisphere.
For $\cep=25^\circ$ ($=205^\circ-180^\circ$) the opposite emission scenario and the dominance of peak \B{A} is observed. This implies that for these CEP-values the two ionization steps each take place during predominantly only one sub-cycle burst: {up-bursts} for $\cep=205^\circ$ and {down-bursts} for $=25^\circ$.

The simulated data for $\cep=205^\circ$ shown in the leftmost column of Fig.~\ref{fig3} confirm this single-burst ionization dynamics and the experimentally observed yield-asymmetry.
For intermediate values of the CEP, e.g. for $\cep \approx 110^\circ$, the numerical model reveals that more than one pair of sub-cycle bursts contribute and both up- and down-pairs are emitted. Accordingly, the measured electron yield in Fig.~\ref{fig2}(a) does not show any asymmetry for $\cep \approx 110^\circ$ and peaks \B{A} and \B{A'} are equally strong.
The two-electron emission dynamics for $\Delta t=2T$ that results in the peaks \B{C/C'} takes place analogously to the case $\Delta t=1T$. The measured CEP-dependence of the corresponding yield-distributions in Fig.~\ref{fig2}(c) can therefore be explained qualitatively using the same argumentation.

Of particular interest is the case $\Delta t = 1.5T$. For this value, the two electrons are streaked into opposite hemispheres and dominantly only peak \B{B} is observed.
The corresponding CEP-resolved momentum distribution in Fig.~\ref{fig2}(b) for $\Delta t = 1.5T(\pm 0.25T)$ shows that the mean value of peak \B{B} oscillates with CEP.
To explain this CEP-oscillation, we turn to the simulated distributions in Fig.~\ref{fig3}. The center column shows
that for $\cep \approx 90^\circ$ dominantly only one pair of sub-cycle bursts is emitted symmetrically around the pulse peak. One electron is streaked upwards, the other one downwards, and their momenta almost cancel, resulting in very small sum momentum close to zero.
For $\cep \approx 180^\circ$ (Fig.~\ref{fig3}, rightmost column), in contrast, each electron is emitted during two sub-cycle peaks and streaked both up- and downwards.
Hence, the photoelectrons are no longer emitted symmetrically around the pulse maximum, explaining the CEP-oscillation of $\vec{p}_{\textrm{rec},\perp}$.
We conclude from this analysis that it is possible to control the SDI process with the CEP such that the two emitted electrons can be disentangled in the experimental distributions, i.e., that a situation as in the center column of Fig.~\ref{fig3} can be achieved.

\begin{figure}[tb]
  \centering
  \includegraphics[width=\columnwidth]{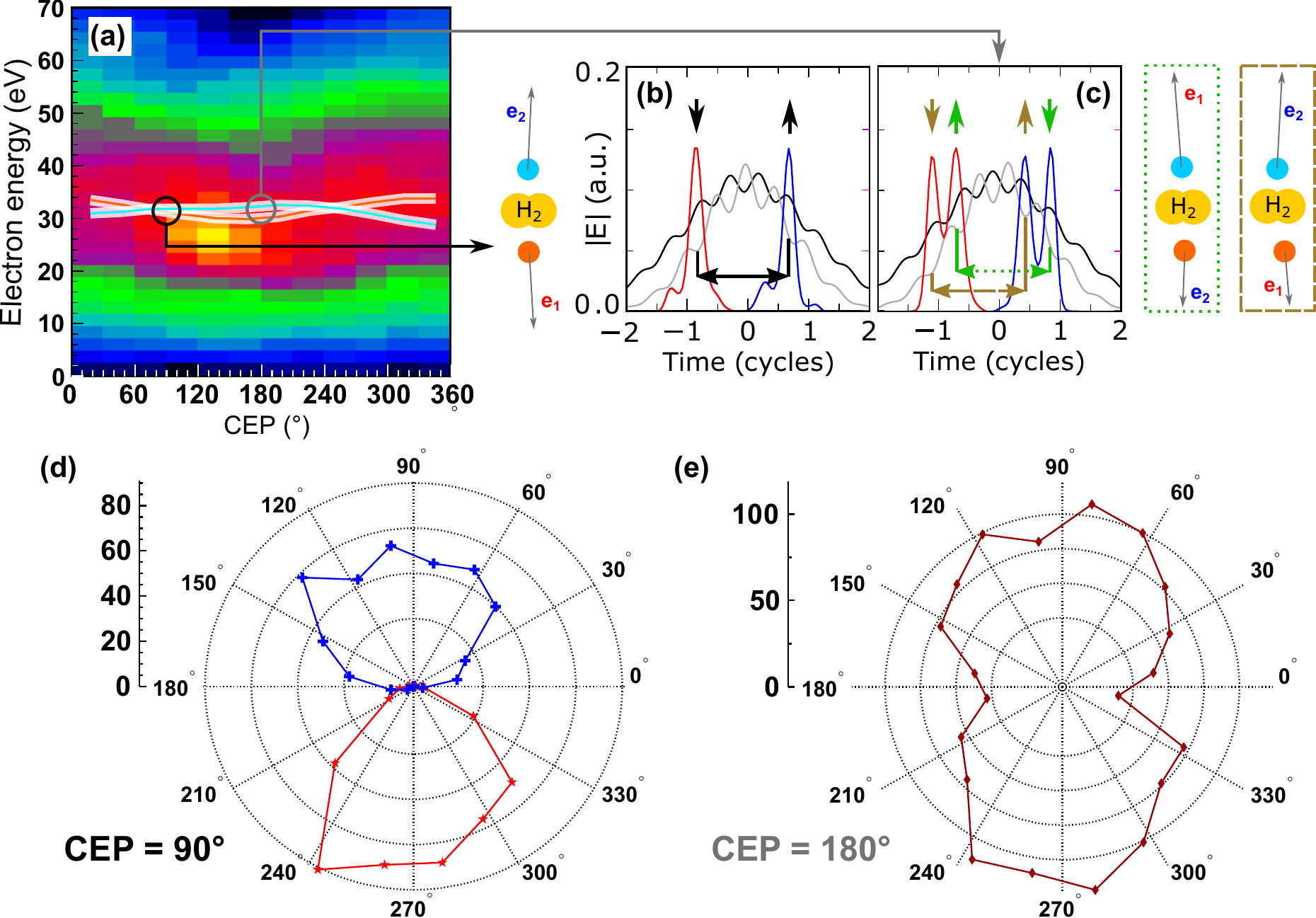}
  \caption{(a) Measured energy distribution of electrons for $\Delta t = 1.5T(\pm 0.25T)$ ($11.6<\textrm{KER}<13.6$\,eV, peak \B{B}) over CEP emitted into the hemisphere facing the detector. The average energy is shown by a red line (white shading indicates experimental uncertainty). The blue line denotes the average energy emitted into the opposite hemisphere. (b,c) Simulated sub-cycle emission bursts (reproduced here for convenience from Fig.~\ref{fig3}) for the same range of $\Delta t$ as in panel (a) for two values of the CEP indicated by circles and arrows. The cartoons left/right of (b)/(c) visualize the corresponding momentum vectors of the two emitted electrons. The two possible cases in (c) are indicated by brown and green colors, respectively. (d) Molecular frame photoelectron angular distribution (MF-PAD) corresponding to (b). (e) MF-PAD corresponding to (c).}
  \label{fig4}
\end{figure}

The value of the CEP for which this is possible can be determined directly from experimental data without the need to consult simulations.
To demonstrate this, we consult the electron energy distribution over CEP shown in Fig.~\ref{fig4}(a). The mean energy of the electrons emitted towards the detector, marked by a red line, shows a clear oscillation with the CEP. If those electrons that are streaked into the opposite hemisphere (away from the detector) are selected, an oscillation phase-shifted by 180$^\circ$ is obtained (blue line). From the discussions above
we know that if each electron is emitted during one sub-cycle burst only, the two electrons will be emitted symmetrically around the pulse peak---one upwards the other downwards. As a consequence, both electrons must have very similar energy. Thus, for symmetry reasons, the CEP at which this situation occurs is that where the energy of the upward and downward electrons is the same. This is, in accordance with the simulations, the case for $\cep = 90^\circ$ (and $\cep = 270^\circ$), indicated by a black circle in Fig.~\ref{fig4}(a). The corresponding emission scenario and the simulated sub-cycle bursts in the time-domain (reproduced from Fig.~\ref{fig3})
are sketched in Fig.~\ref{fig4}(b). 
On the contrary, when the electrons are emitted during more than one pair of sub-cycle bursts, the upwards and downwards electrons will have markedly different energy. This is the case for $\cep = 180^\circ$, indicated by a grey circle in Fig.~\ref{fig4}(a); the corresponding emission scenario
is sketched in Fig.~\ref{fig4}(c).

Fig.~\ref{fig4} thus holds experimental proof that
for $\cep \approx 90^\circ$ and $\cep \approx 270^\circ$
it becomes possible to disentangle the contributions of the sub-cycle bursts in electron momentum distributions simply by selecting one electron from the upwards and the other from the downwards hemisphere. Moreover, based on the known rotation direction of $\vec{E}(t)$ determined by the helicity of the laser field it is even feasible to determine during which of the two sub-cycle bursts a specific electron has been emitted.
For the helicity used in the experiment
and for $\cep = 90^\circ$ the first emitted electron is streaked downwards [cf. Fig.~\ref{fig4}(b)], for $\cep = 270^\circ$ the first emitted electron is streaked upwards. With that, by fixing the CEP to one of the two values, the SDI dynamics is completely determined.



This opens up the possibility to obtain an experimental trace of the attosecond evolution of the sub-cycle ionization bursts.
By exploiting the fact that the electron emission dynamics is streaked into the angular direction by the rotating laser field we can represent the attosecond evolution of the sub-cycle two-electron bursts as photoelectron angular distributions (PADs).
Moreover, because in our experiment we also measure the orientation of the molecules in the lab frame via the momentum vectors of the two protons, we are able to plot the PADs in the molecular frame (MF) of reference. The MF-PADs for $\cep=90^\circ$ and $\cep=180^\circ$ are shown in Fig.~\ref{fig4}(d) and (e).
Because for $\cep = 180^\circ$ two pairs of sub-cycle bursts are emitted, the MF-PAD cannot be separated without ambiguity, cf. Fig.~\ref{fig4}(e).
%
In contrast, the dominant single pair emitted for $\cep=90^\circ$ can be unambiguously disentangled in the MF-PAD, see Fig.~\ref{fig4}(d). 

The capability of obtaining absolutely defined traces of the attosecond two-electron emission for specific molecular orientations in the laser polarization plane opens up intriguing possibilities. For example, it allows to trace the intramolecular electron dynamics in between the two ionization steps \cite{Hennig2005, Takemoto2010, Odenweller2011}.
%
These dynamics can lead to very complicated pattern of the intramolecular charge density \cite{Spanner2012, Hennig2005} that may modulate the sub-cycle ionization behavior. We could, however, show that such localization dynamics is not significant in H$_2^+$ for the small ionization delays $\Delta t \lesssim 2T$ considered here \cite{Hanus2019}.
Our method can also be applied to separately determine the Coulomb-distortion \cite{Eckle2008, Torlina2015} imparted by the singly respectively doubly charged parent ion onto the first respectively second emitted electron.



In conclusion, we observed the sub-cycle ionization bursts in SDI of H$_2$ with elliptically polarized intense few-cycle laser pulses with a known CEP using multi-particle coincidence momentum imaging.
%
This permitted us to measure the MF-PAD for a single pair of sub-cycle bursts leading to double ionization, and the unambiguous assignment of each burst to one of the two detected electrons.
Although demonstrated for H$_2$, we expect that the approach demonstrated here will allow the observation of the sub-cycle ionization bursts also in more complicated molecules, as the only requirement for the molecule is a suitably fast fragmentation dynamics.
This will grant experimental access to the Coulomb distortion for specific charge states and to the intramolecular electron localization dynamics on attosecond time-scales in polyatomic molecules.


\acknowledgments
This work was financed by the Austrian Science Fund (FWF), Grants No.
P28475-N27,  
and P30465-N27.  


%

\end{document}


{\bf \noindent \sffamily \huge \color{titlecolor} Experimental Separation of Sub-Cycle Ionization Bursts in Strong-Field Double Ionization of H$_2$}

\vspace{0.8cm}

{\bf \noindent \sffamily \Large Supplemental Material}

\vspace{0.8cm}

{\bf \noindent \large V\'{a}clav\,Hanus, Sarayoo\,Kangaparambil, Seyedreza\,Larimian, Martin Dorner-Kirchner, Xinhua\,Xie, Markus\,S.\,Sch\"{o}ffler, Gerhard\,G.\,Paulus, Andrius\,Baltu\v{s}ka, Andr\'{e} Staudte, Markus\,Kitzler-Zeiler}


\section{Simulations \label{sec:simulations}}

\subsection{Ionization dynamics and sub-cycle bursts \label{sec:ionization_dynamics}}

To simulate the ionization dynamics and electron momentum distributions shown in, e.g., Fig.~3 in the main manuscript, 
we calculate the double ionization and nuclear dynamics using (semi-)classical models for a laser electric field of the form
\begin{equation}\label{equ:laserfield}
\vec{E}(t) = \vec{E}_\parallel \cos(\omega t + \cep) + \vec{E}_\perp \sin(\omega t + \cep),
\end{equation}
where $\vec{E}_\parallel$ and $\vec{E}_\perp = \varepsilon \vec{E}_\parallel$ denote the peak electric field strengths along the major respectively minor axis of the  polarization ellipse, $\varepsilon$ is the ellipticity, and $\cep$ denotes the carrier-envelope phase.
Because in our experiments the width of the supersonic gas jet along the laser propagation direction (perpendicular to the $\parallel$ and $\perp$ directions) was cut by adjustable razor blades to about 20\,$\mu$m, much shorter than the Rayleigh length of the laser beam ($\approx$200\,$\mu$m), we neglected the dependence of the laser electric field along the laser propagation direction.

Ionization is modeled as described in Ref.~\citen{Tong2004} using a modified ADK-type formula. For the first ionization step, taking place at time $t_1$, the value of the ionization potential is $I_{p,1} = 15.4$ eV \cite{Dibeler1965}. 
Ionization-depletion \cite{Tong2004} is included for the first ionization step.  
The first ionization step triggers nuclear motion in the cation H$_2^+$.
The nuclear motion is modeled classically as described in Section~\ref{sec:nuclear_motion}. 
The second ionization step happens $\Delta t$ after the first ionization step at time $t_2=t_1+\Delta t$, when the two nuclei have reached a certain internuclear distance, $R$. The ionization potential for the second ionization step, $I_{p,2}(R)$, decreases monotonically with $R$. 
We use a linear dependence of the ionization potential $I_{p,2}$ on $R$. The offset and slope of the linear function $I_{p,2}(R)$ are obtained by a best fit of the simulated to the measured distributions of the kinetic energy release (KER) that results upon fragmentation of the doubly charged parent ion, see Fig.~\ref{fig1}(c). The simulated KER distribution is obtained as described in Section~\ref{sec:nuclear_motion}. 

With this model we obtain an ionization rate $w_1(t_1)$ for the first ionization step and an ionization rate $w_2(t_2;R(\Delta t))$ for the second ionization step. 
The double ionization rate for a given value of $\Delta t$ is therefore given by $W(t_1,\Delta t) = w_1(t_1) w_2(t_2=t_1+\Delta t;R(\Delta t))$, where $R(\Delta t)$ is obtained as described in Section~\ref{sec:nuclear_motion}. 
With this, the sub-cycle ionization bursts leading to double ionization for a certain value of $\Delta t$ (e.g. $\Delta t=1T$ or $1.5T$) and specific values of the $\cep$, shown in Figs.~3 and 4 of the main manuscript, as a function of absolute time $t$ within the pulse envelope are obtained by plotting $W(t=t_1,\Delta t)$ for the first ionization step (denoted by red lines) and $W(t=t_2=t_1+\Delta t,\Delta t)$ for the second ionization (denoted by blue lines).

\subsection{Electron momentum distributions \label{sec:electron_mom}}

The momentum distributions of the two emitted electrons shown in Fig.~3 of the main manuscript for specific values of $\Delta t$ and the $\cep$ are calculated by integrating the double ionization rate $W(t_1,\Delta t) = w_1(t_1) w_2(t_2=t_1+\Delta t;R(\Delta t))$, described in Section~\ref{sec:ionization_dynamics}, over $t_1$ and binning the values of $W(t_1,\Delta t)$ on a grid for the two electrons' momenta $\vec{p}_{e1}$ and $\vec{p}_{e2}$. The momenta are obtained from the classical formula $\vec{p}_{ei}=-\vec{A}(t_i), \; i=1,2$ with  $\vec{A}(t)=-\int_{-\infty}^t \vec{E}(t') dt'$ derived from the laser electric field defined in Equ.~(\ref{equ:laserfield}). 

\subsection{Nuclear motion \label{sec:nuclear_motion}}

Nuclear motion is modeled by solving Newton's equations on the \ssg{}  potential energy curve given in Ref.~\citen{Sharp1970}. 
We assume that upon the first ionization step at the instant $t_1$ the \ssg{} curve of H$_2^+$ is populated at an internuclear distance $R$ given by the equilibrium distance of H$_2$. 
We further assume that the potential energy curve is unaffected by the laser electric field and that the nuclear motion is determined purely by the gradient of  the \ssg{} curve at any given internuclear distance.
The second ionization step takes place at an instant $t_2=t_1+\Delta t$. At $t_2$ the repulsive Coulomb curve is populated and the KER (in atomic units) is calculated as $\text{KER}=\frac{1}{R} + E_K$, where $E_K$ denotes the kinetic energy of the protons acquired during their motion prior to the Coulomb explosion.
To obtain a KER distribution from our simulations we loop over the times $t_1$ and $t_2=t_1+\Delta t$ and calculate for each combination the double ionization rate  $W(t_1,\Delta t)$ (see Section~\ref{sec:ionization_dynamics}). By summing up all combinations of $t_1$ and $t_2>t_1$ with their respective double ionization probability $W$ we obtain a distribution of KER, shown in Fig.~\ref{fig1}(c) for the laser pulse parameters used in the experiments, in comparison with the measured KER distribution.


\section{CEP-dependent asymmetry of photoelectron emission}

\begin{figure*}[tb]
  \centering
  \includegraphics[width=0.8\textwidth]{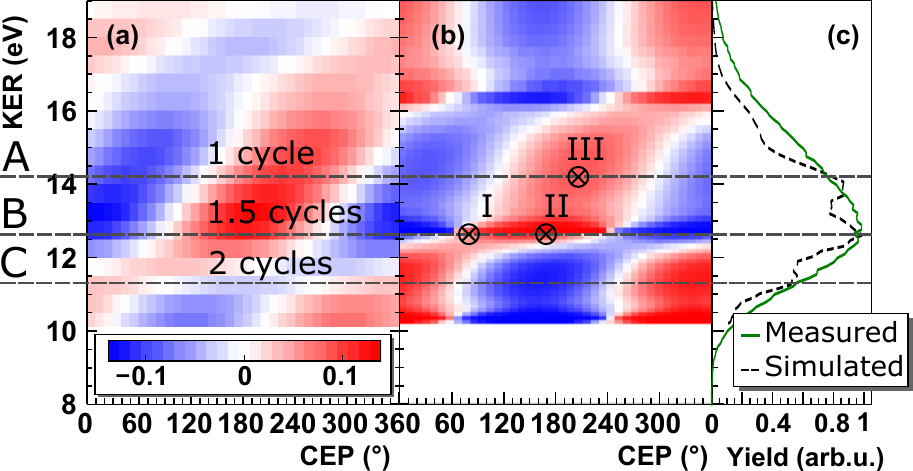}
  \caption{Measured (a) and simulated (b) asymmetry $A$ defined in Equ.~(\ref{eq:asymmetry}) as a function of KER and $\cep$. (c) Measured distribution of KER in comparison with the simulated one obtained as described in Section~\ref{sec:nuclear_motion}.
  KER values corresponding to an ionization delay of 1, 1.5 and 2 laser cycles, marked by  \textbf{A,B,C} as in the main article, are indicated. Three specific points in the $A$-distribution, marked I, II and III are indicated, see text for details.}
  \label{fig1}
\end{figure*}

As shown in Fig.~2 of the main manuscript and described in text corresponding to this figure, the two-electron emission dynamics strongly depends on the $\cep$. For certain values of the $\cep$ most of the electrons are emitted with 
$\vec{p}_{\textrm{rec},\perp} < 0$ (or $\vec{p}_{\textrm{rec},\perp} > 0$).
This asymmetric $\cep$-dependent electron emission can be quantified using the dimensionless asymmetry parameter 
\begin{equation}
\label{eq:asymmetry}
A = \frac{N(\vec{p}_{\textrm{rec},\perp}<0) - N(\vec{p}_{\textrm{rec},\perp} > 0)}{N},
\end{equation}
where $N(\vec{p}_{\textrm{rec},\perp} < 0)$ and $N(\vec{p}_{\textrm{rec},\perp} > 0)$ denote the number of ionization events that resulted in a negative respectively positive recoil momentum along the direction perpendicular to the main axis of the polarization ellipse, $\vec{p}_{\textrm{rec},\perp} = - (\vec{p}_{e1,\perp} + \vec{p}_{e2,\perp})$, and $N$ the total number of ionization events.   
The measured asymmetry parameter $A(\textrm{KER},\cep)$ as a function of KER and $\cep$ is shown in Fig.~\ref{fig1} in comparison with the simulated asymmetry distribution obtained with the model described in Section~\ref{sec:simulations} for the laser field defined in Equ.~(\ref{equ:laserfield}). 
For the comparison in Fig.~\ref{fig1} the constant offset value of the $\cep$ of the experimental data, which is inaccessible in the stereo-ATI technique \cite{Rathje2012,Sayler2011a} used in our experiments for determining the CEP,  was shifted to obtain best agreement with the simulations. This calibration of the $\cep$-offset was used throughout this document and also the main article.

The dependence of the asymmetry $A$ on $\cep$ for the three specific cases where $\Delta t$ is 1, 1.5 and 2 laser cycles is described in detail in the main article in the text corresponding to Figs.~2 and 3. 
There, it is explained that the $\cep$ can determine whether both electrons are emitted symmetrically around the pulse peak, resulting in $A\approx 0$ (point I in Fig.~\ref{fig1}), or whether they are emitted asymmetrically, resulting, e.g., in a pronounced value  $A> 0$ (point II in Fig.~\ref{fig1}).
What is not visible in the figures in the main article is the dependence of the asymmetry on the KER. Fig.~\ref{fig1} shows that the asymmetry $A$ markedly depends in KER. As can be seen, it is not possible for a given value of the $\cep$ to obtain the same value of   $A$ for all values of KER. 
For different values of KER 
the $\cep$ needs to be adapted to obtain the same value of $A$. For example, to move from point II to point III in the asymmetry distribution, the $\cep$ needs to be shifted by about 40$^\circ$ to stay on the same large value of $A>0$, resulting in the right-tilt of the colored maxima/minima stripes of the asymmetry distribution. 
The reason for this correlation between CEP and KER in $A$ is the subtle dependence of the electron momenta on the temporal structure of the sub-cycle ionization bursts discussed in the main manuscript in connection with Fig.~3. From this discussion it becomes clear that the bursts' temporal structure depends differently on the CEP for every value of $\Delta t$. 
Therefore, since $\Delta t$ is monotonically connected to the value of KER, as explained in Section~\ref{sec:simulations}, the CEP is connected in the same way to KER, which explains the correlation between CEP and KER in Fig.~\ref{fig1}.

\hspace{1cm}

\bibliography{refs_markus}